\documentstyle[epsf,psfig,multicol,aps,prb,tighten]{revtex}
\renewcommand{\narrowtext}{\begin{multicols}{2}
\global\columnwidth20.5pc}
\renewcommand{\widetext}{\end{multicols}
\global\columnwidth42.5pc}
\def\top#1{\vskip #1\begin{picture}(290,80)(80,500)\thinlines \put(
65,500){\line( 1, 0){255}}\put(320,500){\line( 0, 1){
5}}\end{picture}}
\def\bottom#1{\vskip #1\begin{picture}(290,80)(80,500)\thinlines \put(
330,500){\line( 1, 0){255}}\put(330,500){\line( 0, -1){
5}}\end{picture}}
\newcommand{\pprl}{Phys. Rev. Lett. \ }
\newcommand{\pprb}{Phys. Rev. {B}}
\newcommand{\eq}{\begin{equation}}
\newcommand{\ee}{\end{equation}}
\newcommand{\eqa}{\begin{eqnarray}}
\newcommand{\eea}{\end{eqnarray}}

\begin{document}
\draft
\title{Spectral Properties of a Quantum Impurity
in d-Wave Superconductors}

\author{Xi Dai and Ziqiang Wang}
\address{Department of Physics, Boston College,
Chestnut Hill, MA02467}
\date{\today}
\maketitle

\begin{abstract}
The variational approach of Gunnarsson and Sch\"onhammer to the 
Anderson impurity model is generalized to study d-wave superconductors 
in the presence of dilute spin-1/2 impurities. We show that the local moment 
is screened when the hybridization exceeds a nonzero critical value at which 
the ground state changes from a spin doublet to a spin singlet. The electron 
spectral functions are calculated in both phases. We find that while a 
Kondo resonance develops above the Fermi level in the singlet phase, 
the spectral function exhibits a low-energy spectral peak below the 
Fermi level in the spin doublet phase. The origin of such a 
``virtual Kondo resonance'' is the existence of low-lying collective 
excitations in the spin-singlet sector. We discuss our results in connection 
to recent spectroscopic experiments in Zn doped high-T$_c$ superconductors.
\end{abstract}
\narrowtext

The physics of quantum impurities in high-T$_c$ superconductors has generated
an increasing amount of experimental and theoretical interests. 
NMR measurements in YBCO showed that replacing the Cu-atom in the CuO$_2$ plane
by a nonmagnetic Zn or Li atom induces a spin-1/2 magnetic
moment on the Cu ions in the vicinity of the impurity
\cite{julien00,bobroff01}. Atomic resolution scanning tunneling
microscope (STM) experiments on BSCCO
found that the local density of states (DOS) near the Zn impurity
exhibits a sharp resonance peak just below the Fermi energy \cite{pan}. 
The origin of the resonance peak has been attributed 
to potential scattering \cite{salkola,flatte}, 
local moment formation \cite{wanglee}, and magnetic Kondo 
scattering \cite{sachdev,vojta,zhu,zhang} where it is identified with the 
Kondo resonance arising from the screening of the local moment by the
electrons in d-wave superconductors (dSC).
While it remains controversial as to which of the above is the
dominant mechanism responsible for the low-energy conductance peak,
and how the findings of NMR are reconciled with those of STM, it has
become important to acquire basic understandings of the spectral 
properties of quantum impurities in dSC.

The ground state and the impurity spectral function in the
{\it non-superconducting} phase with a vanishing density of states (DOS), 
i.e. the pseudogap phase, have been studied in detail
following the work of Withoff and Fradkin \cite{fradkin,vojta}.
In this paper, we study these properties associated with
the Kondo physics in the d-wave
{\it superconducting} phase. In addition to the ground state
properties, we determine the momentum dependent, conduction electron 
spectral function $A(k,\omega)$ in dSC.
This is an important quantity since it is directly
measured by angle-resolved photoemission spectroscopy (ARPES).
To this end, we describe the localized impurity by the
infinite-U Anderson model and generalize
the variational large-N approach of Gunnarsson and Sch\"onhammer
\cite{gs,book} for rare earth and actinide heavy fermion materials 
to the case of dSC. This is a reliable, physically transparent
many body approach that allows an essentially analytical
calculation of $A(k,\omega)$.
Specifically, starting from the singlet BCS state with
the impurity site empty (singlet) or singly occupied (doublet),
we construct the variational wave functions of states in both the spin singlet
and doublet sectors from a basis of states generated by the hybridization
between the impurity and the conduction
electrons to next to leading order in $1/N$.
We find that there exists a critical value for the hybridization 
that separates a spin doublet ground state from
a singlet ground state where the local moment is fully screened.
Using the variational wave functions, we calculate the T-matrix, the
self-energy, and the spectral function of the conduction electrons in
both the singlet phase and the local moment phase.
We find that while a Kondo resonance naturally develops above the Fermi level 
in the singlet phase, $A(k,\omega)$ exhibits a spectral
peak of a similar strength below the Fermi energy in the local moment phase.
The origin of this {\it virtual Kondo resonance} can be traced to
the existence of low-lying collective spin singlet excitations
above the doublet ground state in the local moment phase.
The qualitatively different behaviors of the spectral function
make it possible to distinguish between the two ground
states by STM and ARPES.

The single impurity Anderson model in a dSC
can be written as $H=H_c+H_d+H_v$,
\eqa
H_c&=&\sum_{k\sigma} \epsilon_k c^{\dagger}_{k\sigma}c_{k\sigma} + \sum_{k}
\Delta_k (c_{k\uparrow}^{\dagger}
c_{-k\downarrow}^\dagger + h.c.)
\label{hc0} \\
H_d&=& \epsilon_d\sum_{\sigma}d^{\dagger}_{\sigma}d_{\sigma}
+Ud^{\dagger}_{\uparrow}d_{\uparrow}d^{\dagger}_{\downarrow}d_{\downarrow}
\label{hd0} \\\
H_v&=& V\sum_{k\sigma} ( c^{\dagger}_{k\sigma}d_{\sigma}+ h.c.).
\label{hv0}
\eea
Here $c_{k\sigma}^\dagger$ creates an electron with a dispersion
$\epsilon_k=-2t(\cos k_x+\cos k_y)-\mu$ in a dSC
described by the gap function $\Delta_k={1\over 2}\Delta_0(\cos k_x-\cos k_y)$;
$d_\sigma^\dagger$ creates a localized electron with
energy $\epsilon_d$ at the impurity site, and $U$ is the on-site Coulomb
repulsion. We take the infinite-U limit such that the impurity site
is either empty or singly occupied.
The hybridization between the impurity and the conduction electrons
is described by the hybridization constant $V$ in $H_v$.
For simplicity, only the case of $N=2$ is written
explicitly in Eq.~(\ref{hc0}-\ref{hv0}).
The generalization to the case of $N$ orbital degeneracy
with $NV^2={\cal O}(1)$ is straightforward \cite{gs,book,spn}.

As in the BCS theory, $H_c$ can be diagonalized by a Bogoliubov transformation:
$
c^{\dagger}_{k\uparrow}=u_k\alpha^\dagger_{k\uparrow}-v_{k}
\alpha_{-k\downarrow},
$
and
$
c^{\dagger}_{-k\downarrow}=u_k\alpha^\dagger_{-k\downarrow}
+v_{k}\alpha_{k\uparrow},
$
where $u_k^2=1-v_k^2=(1+\epsilon_k/E_k)/2$
with
$
E_k=\sqrt{\epsilon_k^2+\Delta_k^2}.
$
$H_c$ and $H_v$ can be expressed in terms of the quasiparticles,
$
H_c=\sum_{k\sigma}E_k\alpha_{k\sigma}^\dagger\alpha_{k\sigma},
$
and
\eqa
H_v&=&V \sum_{k}\bigl[(
u_k\alpha^\dagger_{k\uparrow}-v_{k}\alpha_{-k\downarrow})
d_{\uparrow}
\nonumber \\
&+&(u_{k}\alpha^\dagger_{-k\downarrow}+v_{k}\alpha_{k\uparrow})d_{\downarrow}
+h.c.\bigr].
\label{hv}
\eea
%
%
\begin{figure}
\center
\centerline{\epsfxsize=3.2in
\epsfbox{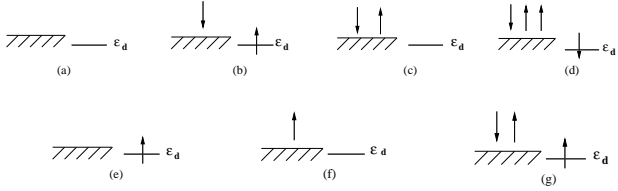}}
\begin{minipage}[t]{8.1cm}
\caption{Schematics of the variational states. Singlet (first row) and
doublet (second row) states to leading (a,b, and e) and next to leading
order (c,d,f, and g) in $1/N$.}
\label{fig1}
\end{minipage}
\end{figure}

We now generalize the variational method of Gunnarsson and Sch\"onhammer
to the case of superconductors \cite{arovas}.
Since the BCS ground state $\vert BCS\rangle$ with the impurity level empty
is a singlet (Fig.~1a), we can construct
other states in the singlet sector by acting on $|BCS\rangle$ with
the Hamiltonian $H$. Operating once with the hybridization term
generates the type of states $\vert \epsilon_d {\bar\sigma}E_k\sigma\rangle$
shown in Fig.~1b in which a quasiparticle
excitation with spin $\sigma$ is created from the dSC condensate and
the impurity level is singly occupied by an electron with the opposite
spin ${\bar\sigma}$.
Taking the linear combination of all such states gives the
trial wavefunction for the singlet state:
\eqa
|S\rangle &=&A^s\bigl[|BCS\rangle+{1\over {\sqrt{N}}} \sum_{k_1\sigma}
B_{k_1}^s |\epsilon_d {\bar\sigma} E_{k_1} \sigma\rangle
\nonumber \\
&+&\sum_{k_1k_2\sigma} {1\over {\sqrt{N}}} C_{k_1k_2}^s
|E_{k_1}{\bar\sigma} E_{k_2}\sigma\rangle
\nonumber \\
&+&\sum_{k_1 k_2 k_3 \sigma \sigma^\prime} {1\over {N}}
D_{k_1k_2k_3}^s |\epsilon_d {\bar\sigma}^\prime
E_{k_1}{\bar\sigma} E_{k_2}\sigma E_{k_3}\sigma^\prime\rangle\bigr].
\label{s}
\eea
The last two terms in Eq.~(\ref{s}) contain singlet states generated to
the next to leading order in the $1/N$ expansion shown in Figs.~1c and 1d.
They correspond to states with two quasiparticle excitations and an
unoccupied impurity level and three quasiparticles plus a
singly occupied impurity level respectively.
The parameters $A^s$ to $D^s$ are determined by minimizing the energy
$E_s=\langle S\vert H\vert S\rangle/\langle S\vert S\rangle$. We find
\eqa
B_{k}^s&=&\sqrt{N}V v_k[E_s-\epsilon_d-E_k-\Sigma_1(E_s-E_k)]^{-1},
\nonumber \\
C_{k_1k_2}^s&=&
Vu_{k_2}B_{k_1}^s[ E_s-E_{12}-\Sigma_0(E_s-\epsilon_d
-E_{12})]^{-1},
\nonumber \\
D_{k_1k_2k_3}^s&=&\sqrt{N}Vv_{k_3}C_{k_1k_2}^s
[E_s-\epsilon_d-E_{123}]^{-1},
\label{paras}
\eea
where $E_{12}=E_{k1}+E_{k2}$, $E_{123}=E_{k_1}+E_{k_2}+E_{k3}$,
and 
\eqa
E_s&=&\sum_k{ NV^2v_k^2\over E_s-\epsilon_d-E_k-\Sigma_1(E_s-E_k)},
\label{es} \\
\Sigma_0(\omega)&=&NV^2\sum_k{v_k^2}(\omega-E_k)^{-1},
\label{sigma0}\\
\Sigma_1(\omega)&=&V^2\sum_k{u_k^2}
[\omega-E_k-\Sigma_0(\omega-\epsilon_d-E_k)]^{-1}.
\label{sigma1}
\eea
\begin{figure}
\vspace{-0.5truecm}
\center
\centerline{\epsfysize=2.6in
\epsfbox{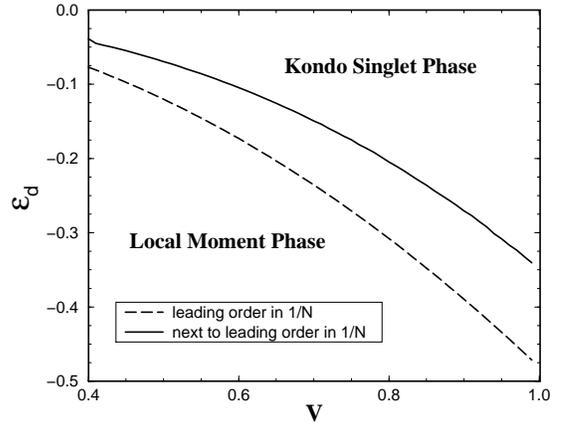}}
\vspace{-0.5truecm}
\begin{minipage}[t]{8.1cm}
\caption{Phase diagram of the infinite-U Anderson model in a dSC
to leading (dashed line) and next to leading order
(solid line) in 1/N.}
\label{fig2}
\end{minipage}
\end{figure}

The variational state in the spin doublet sector
can be constructed starting from the BCS state plus a singly occupied
impurity level (Fig.~1e) in the large-N limit.
To next to leading order in $1/N$,
two types of states are generated: one corresponds to a quasiparticle
excitation and an empty impurity level (Fig.~1f) and the other to two
quasiparticles out of the condensate and a singly occupied impurity
level shown in Fig.~1g. Denoting the net moment in the doublet
state by $\tau$, we have
\eqa
|D\tau\rangle&=&A^d\bigl[|BCS \epsilon_{d}\tau\rangle+\sum_k
B_k^d |E_{k}\tau\rangle
\nonumber \\
&+&{1\over\sqrt{N}}\sum_{k_1k_2\sigma}C_{k_1k_2}^d
|\epsilon_{d}{\bar\sigma} E_{k_1}\tau E_{k_2}\sigma\rangle\bigr].
\label{d}
\eea
As in the singlet case, the variational parameters
$A^d$, $B^d$, $C^d$, and the
total energy are determined by minimizing
$E_d=\langle D\tau\vert H\vert D\tau\rangle/\langle D\tau\vert D\tau\rangle$
in the doublet state:
\eqa
B_{k}^d&=&Vu_k [E_d-E_{k}-\Sigma_0(E_d-E_{k}-\epsilon_d)]^{-1},
\nonumber \\
C_{k_1k_2}^d&=&\sqrt{N}V v_{k_2}B_{k_1}(E_d-E_{k_1}-E_{k_2}-\epsilon_d)^{-1},
\nonumber \\
E_d=\epsilon_d&+&\sum_k {V^2u_k^2\over
E_d-E_{k}-\Sigma_0(E_d-E_{k}-\epsilon_d)}.
\label{ed}
\eea

The ground state and the phase diagram are obtained
by comparing $E_s$ and $E_d$ derived in Eqs.~(\ref{es})
and (\ref{ed}) as a function of the impurity level
$\epsilon_d$ and the hybridization $V$.
We use the d-wave gap $\Delta_0$ as
the energy unit, set $t$ to $\Delta_0$, and integrate over
$k$ in the continuum limit. We find that, due to the linearly
vanishing DOS in the dSC, the impurity
spin can be fully screened by the Bogoliubov quasiparticles only
if the hybridization $V$ is greater than a nonzero value.
Similar to the pseudogap Kondo impurity model \cite{vojta,fradkin},
both the spin singlet and the doublet local
moment phases exist for an Anderson impurity in a dSC
and are separated by a line of phase transitions in the phase 
diagram shown in Fig.~2.

One of the advantages of the variational wave function approach
is that it allows a reliable and transparent many-body calculation 
of the spectral function \cite{gs},
$A(k,\omega)=(-1/2\pi){\rm Im}\overline{G_c}(k,\omega)$, where
$\overline{G_c}(k,\omega)$ is the impurity averaged, retarded
normal Green's function for the conduction electrons. 
The local Green's function
$G_c(i,j,\tau,t)=-i\theta(t)\langle\phi_0\vert\{c_{i\sigma}(t),
c_{j\sigma}^\dagger(0)\}\vert\phi_0\rangle$ where
$\vert\phi_0\rangle=\vert S\rangle,\vert D\tau \rangle$ represent
the fully interacting singlet and doublet (with polarization $\tau$) 
ground states.
Both the local electron Green's function at the impurity site and 
the impurity averaged conduction electron Green's function in momentum space
can be written
in terms of the T-matrix, which can be obtained from the variational
wave functions. Using the Nambu representation, the local Green's function
can be written as:
\widetext
\top{-2.8cm}
\begin{equation}
\widehat G_c({\bf r},{\bf r^\prime},\omega)=\widehat G_c^0
({\bf r}-{\bf r^\prime},
\omega)
+\widehat G_c^0({\bf r},\omega)\widehat T(\omega)\widehat 
G_c^0(-{\bf r^\prime},\omega)
\label{G_local}
%
\qquad \widehat T(\omega)=|V|^2\widehat G_d(\omega),
\end{equation}
where $\widehat G_d(\omega)$ is the local electron Green's function
at the impurity site. The impurity averaged Green's function is given by
\begin{equation}
{\overline {\widehat G_c(k,\omega)}}=\left 
[\omega-\epsilon_k{\hat\sigma}_z-\Delta_k{\hat\sigma}_x-\rho_{\rm imp}
\widehat T(\omega)\right ]^{-1}.
\label{G_ARPES}
\end{equation}
where $\rho_{\rm imp}$ is the density of impurities.
Using the variational wave functions, $G_d(\omega)$ can be 
derived in closed form to first order in $1/N$. 
For the singlet ground state we obtain,
\eqa
G_{d\uparrow}^S(\omega+i\delta)&=&
\langle S|d_{\uparrow}{1\over{\omega+i\delta-H+E_S}}
d_{\uparrow}^\dagger|S\rangle
+\langle S|d_{\uparrow}^\dagger{1\over{\omega+i\delta+H-E_S}}
d_{\uparrow}|S\rangle\approx \nonumber \\
&&A^2 \biggl[ {1\over{\omega-\epsilon_d-\Sigma_d(\omega+i\delta+E_S)+E_S+i\delta}}
+\left ( \sum_{k}{1\over N}B_{k}^s{{u_kV}\over{\omega+E_k-E_S+i\delta}}\right )^2\times\nonumber \\
&&\left ( {1\over{\omega+\epsilon_d+\Sigma_d(-\omega-i\delta+E_S)-E_S+i\delta}}\right )
+\sum_k{1\over N}{B_{k}^s}^2{1\over{\omega+E_k-E_S+i\delta}} \biggr]
\label{G_dS}
\eea
where
$
\Sigma_d(\omega+i\delta)=\sum_k{u^2_k V^2}/({\omega-E_k+i\delta}).
$
In the local moment phase, the spin $SU(2)$ symmetry is broken and
the STM measures the spin averaged tunneling density of states.
Since the total Hamiltonian is invariant when the spin up and down
electrons are interchanged, averaging over the
conduction electron spin is equivalent to averaging over the
polarization of the local moment ground state. We thus obtain
\eqa
&&G_{d\uparrow}^{D}(\omega+i\delta)
={1\over 2}G_{d\uparrow}^{D\uparrow}(\omega+i\delta)+
{1\over 2}G_{d\uparrow}^{D\downarrow}(\omega+i\delta)=\nonumber \\
&&\sum_{\tau=\uparrow,\downarrow}{1\over 2}\left (\langle D\tau|d_{\uparrow}
{1\over{\omega+i\delta-H+E_D}}d_{\uparrow}^\dagger|D\tau\rangle
+\langle D\tau|d_{\uparrow}^\dagger{1\over{\omega+i\delta+H-E_D}}
d_{\uparrow}|D\tau\rangle\right )
\approx \nonumber \\
&&{A^2\over 2} \biggl[ {1\over{\omega+\Sigma_0(-\omega-i\delta-\epsilon_d+E_D)-E_D+i\delta}}
+\left ( \sum_{k}{1\over N}B_{k}^d{{v_kV}\over{\omega-\epsilon_d-E_k+E_D+i\delta}}
\right )^2\times\nonumber \\
&&\left ( {1\over{\omega-\Sigma_0(\omega+i\delta-\epsilon_d+E_D)+E_D+i\delta}}\right )
+\sum_k{2\over N}{B_{k}^d}^2{1\over{\omega-\epsilon_d-E_k+E_D+i\delta}} 
\biggr]
\label{G_dD}
\eea
\bottom{-2.7cm}
\narrowtext

We next present results for the spectral function of the conduction
electrons. The impurity level is fixed at $\epsilon_d=-1.4\Delta_0$.
We choose $V=1.6\Delta_0$ for the doublet local moment phase at a
corresponding valence $\langle n_d\rangle=0.93$, and $V=1.8\Delta_0$
for the singlet state with $\langle n_d\rangle=0.58$.
We will show that along nodal directions of the d-wave gap, where
SC coherence is absent ($u_k^2=1$, $v_k^2=0$ for $\vert k\vert < k_f$), 
the spectral function
exhibits qualitatively different behaviors in the two different phases.
In Figs.~3a and 3b, the solid lines depict the low-energy spectra along the
nodes at $k_x=k_y=0.63 k_f$. In the singlet phase (Fig.~3b), a resonance
peak appears {\it above} the Fermi level, which
is the expected Kondo resonance. Physically, this can be understood
by considering an intermediate state in the infinite-U limit with one less
electron than the singlet ground state. An added electron at
the resonance energy creates a quasiparticle excitation and forms a
singlet state that has a large overlap with the Kondo singlet state.
The intrinsic width of
the Kondo resonance is captured by the present variational wave
function approach. In contrast, the spectrum in the local moment
phase (Fig.~3a) shows, remarkably, a resonance peak {\it below}
the Fermi level.
This peak does not have the usual meaning of a Kondo resonance.
Instead, it can be understood as a {\it virtual Kondo resonance}.
Although the ground state is a doublet and the local moment is
unscreened, the collective singlet states of the type
given by Eq.~(\ref{s}) exist as low energy excitations above
the doublet ground state. When an electron is removed from
the SC condensate, quasiparticle excitations
with the opposite spin can be created to screen the local moment
and occupy the virtual spin singlet state. Note that,in general, 
depending on the sign of the particle-hole asymmetry, the Kondo resonance 
in the singlet phase can be either above (the present case for the 
infinite-U Anderson model) or below the Fermi level. 
The virtual Kondo resonace in the local moment phase 
is located on the opposite side of the Fermi level compared to
the Kondo resonance.
\begin{figure}
\vspace{-0.5truecm}
\center
\centerline{\epsfysize=2.0in
\epsfbox{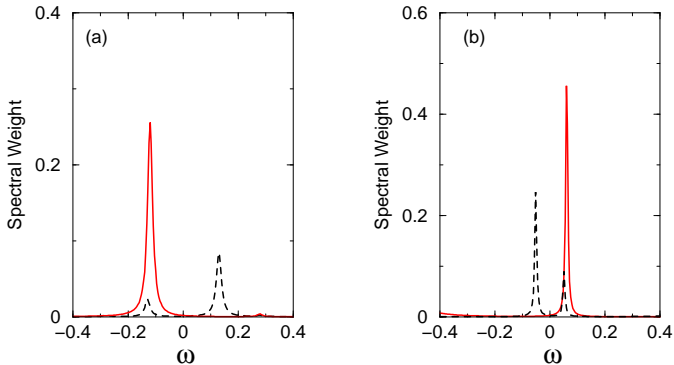}}
\vspace{-0.5truecm}
\begin{minipage}[t]{8.1cm}
\caption{Momentum resolved spectral functions in the local moment (a)
and the singlet (b) phases along the nodal (solid lines) and anti-nodal
(dashed lines) directions.}
\label{fig3}
\end{minipage}
\end{figure}

The behavior of the spectra along anti-nodal directions of
the d-wave gap is shown in dashed lines in Figs.~3a and 3b
at $k_x=k_f$, $k_y=0$. Due to the strong SC coherence
($u_k^2,v_k^2\neq0$), the Kondo and the virtual Kondo resonances are
reflected on the hole and the particle sides such that
the resulting spectra show nearly symmetric resonance peaks.
These results suggest that both ARPES and STM experiments STM can be used to
distinguish between these two ground states.

Finally, we turn to the local density of states (LDOS)
as measured by STM experiments,
$$
\rho(r,\tau,\omega)=-{1\over \pi}\sum_{k,k^\prime,\sigma}
{\rm Im}
G_c(k^\prime\sigma,k\sigma,\tau,\omega)e^{-i(\vec k-\vec {k^\prime})\vec r}.
$$
In Fig.~4, the LDOS at the impurity site is shown for both
the singlet and the local moment
phases. Due to the high symmetry of this tunneling
point, the LDOS spectrum is dominated by the contribution from the low
energy quasiparticle excitations along the nodal directions. As a
result, the tunneling spectrum closely resembles that of 
$A(k,\omega)$ along nodal directions: a single
resonance peak above the Fermi level in the singlet
phase and below the Fermi level in the local moment phase.
The STM data \cite{pan} interpreted this way would suggest that
the Zn-doped BSCCO surface is in the local moment phase.
The LDOS away from the impurity is shown in the insets of Fig.~4.
The conductance spectra in both phases still
exhibit a single peak along nodal directions,
whereas along antinodal directions two peaks appear
due to the SC coherence. The amplitude
of the peaks is significantly reduced than at the impurity site.
\begin{figure}
\center
\centerline{\epsfysize=2.8in
\epsfbox{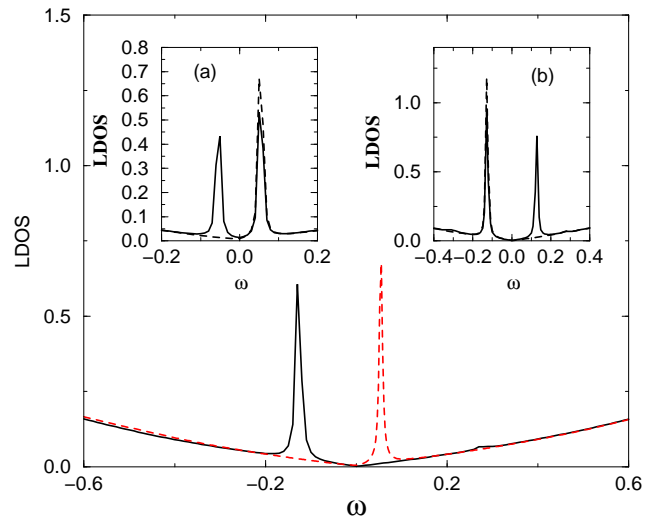}}
\vspace{-0.5truecm}
\begin{minipage}[t]{8.1cm}
\caption{The LDOS spectrum at the impurity site in the singlet (dashed line) 
and the local moment (solid line) phases. The insets show
the LDOS away from the impurity along nodal (dashed lines)
and anti-nodal (solid lines) directions in local moment 
(b) and singlet (a) phases.}
\label{Fig.4}
\end{minipage}
\end{figure}
To conclude, we have studied the spectral properties of electrons in dSC 
coupled to localized Anderson impurities in both the singlet and
the local moment phases using the variational wave function approach.
Our results show that Kondo screening is not a prerequisite
for the emergence of low energy resonance peaks in the spectral function.
Even if the impurity spin is not screened, the virtual Kondo resonance due 
to collective spin singlet excitations would still lead to a sharp 
resonance peak, but located on the opposite side of the Fermi level 
compared to the Kondo resonance in the singlet phase.  
It is the evolution of the resonance peak positions that reveals 
the screening properties of the local moment in the ground states,
which can in principle be measured by ARPES and STM experiments.

The authors thank H. Ding for sharing the ARPES data prior to
publication. This work was supported by DOE DE-FG02-99ER45747,
DE-FG02-02ER63404, and by ACS Petroleum Research Fund. 

\end{multicols}
\end{document}